# The *INTEGRAL* Complete Sample of Type 1 AGN

M. Molina,[1] L. Bassani,[2] A. Malizia,[2] J.B. Stephen,[2] A.J. Bird,[3] A.J. Dean,[3] F. Panessa,[4] A. De Rosa,[4] R. Landi[2]

[1]*IASF/INAF, Via Bassini 15, 20122 Milan, Italy,*
[2]*IASF/INAF, via Gobetti 101, I-40129 Bologna, Italy,*
[3]*School of Physics and Astronomy, University of Southampton, SO17 1BJ, Southampton, U.K.,*
[4]*IASF/INAF, via Fosso del Cavaliere 100, I-00133 Rome, Italy*

**ABSTRACT**

In this paper we discuss the broad-band X-ray characteristics of a complete sample of 36 type 1 AGN, detected by *INTEGRAL* in the 20-40 keV band above the 5.5 $\sigma$ level. We present, for all the objects in the sample, the broad-band (1-110 keV) spectral analysis obtained by using *INTEGRAL*/*Swift*/BAT observations together with *XMM-Newton*, *Chandra*, *ASCA* and *Swift*/XRT data. We also present the general average properties of the sample, i.e. the distribution of photon indices, high energy cut-offs, reflection fractions and absorption properties, together with an in-depth analysis of their parameter space. We find that the average Seyfert 1 power law has an index of 1.7 with a dispersion of 0.2. The mean cut-off energy is at around 100 keV, with most objects displaying $E_c$ in the range 50-150 keV; the average amount of Compton reflection is 1.5 with a typical dispersion of 0.7. We do not find any convincing correlation between the various parameters, an indication that our analysis is not strongly dependent by the interplay between them. Finally, we investigate how the results presented in this work fit into current frameworks for AGN spectral modeling and Cosmic Diffuse X-ray Background synthesis models.

## 1 INTRODUCTION

It is now widely accepted that the Cosmic X-ray Background (CXB) is the product of the integrated emission of point-like extragalactic sources, most of them now resolved below 10 keV thanks to the survey studies conducted both by *XMM-Newton* and *Chandra*. It has also been shown that most of the sources contributing to the CXB are AGN, both unobscured and obscured (e.g. Barger et al. 2005), as already predicted by AGN synthesis models (Comastri et al. 1995). However, while below 10 keV the shape and the intensity of the CXB is quite well understood (Hickox & Markevitch 2006), at higher energies, around 30 keV or above, these quantities are less well studied and hence the AGN contribution is less defined. Recent analyses of *BeppoSAX*, *INTEGRAL* and *Swift* data have allowed a better estimate of the CXB spectrum above 10 keV (see Ajello et al. 2008 and references therein). All these measurements agree with an early HEAO A1 result (Gruber et al. 1999) and are consistent within their systematic uncertainties. Such consistency is not present with data at lower energies, where the measured spectrum may be up to ∼40% larger. Synthesis models need now to take this discrepancy into consideration.

According to AGN synthesis models (e.g. Comastri et al. 2006), several parameters must be taken into account to reproduce the shape of the CXB above 30 keV. These are the covering fraction and the geometry of the cold dense gas responsible for the reflection hump seen around 30 keV; the fraction of highly obscured AGN (Compton Thick sources with $N_H \gtrsim 10^{24} cm^{-2}$); the high energy cut-off observed in the primary continuum emission, the exact value of which is still not well determined. Some recent models (Gilli et al. 2007; Gandhi et al. 2007) also propose the average power law photon index (and its spread in values) and the reflection component as other important input information that needs to be considered when estimating the AGN contribution to the CXB.

It is therefore clear that the determination of all these parameters over wide samples of sources, and particularly over a wide range of energies (above all up to 100 keV and beyond), is important in order to obtain a firm description of the AGN contribution to the CXB and hence a better understanding of the accretion history of the Universe.

The determination of the slope of the continuum emission of AGN and its high energy cut-off, besides being a key element in CXB synthesis models, is also essential for spectral modeling of AGN, since these two parameters are deeply linked to the physical characteristics of the Comptonising region around the central nucleus. Up to now, modeling of the high energy spectra of AGN has focused on how to reproduce and explain the observed shape of the primary continuum. A good fraction of the proposed mod-els ascribe the power law to the inverse Compton scattering of soft photons in a bath of "hot electrons" (e.g. Maraschi & Haardt 1997; Zdziarski et al. 1998). Variations to these basic models depend on the energy distribution of the electrons and their location in relation to the accretion disc.

Measuring both the primary continuum and its cut-off energy is therefore crucial for understanding models and discriminating between them. While the photon index distribution has been well investigated (e.g. Reeves & Turner 2000; Piconcelli et al. 2005), observational results on the cut-off energy have so far been limited by the scarcity of measurements above 10-20 keV, with most of the information coming from broad-band spectra provided by the *BeppoSAX* satellite. Analysis of type 1 and 2 AGN (Perola et al. 2002, Dadina 2007) gives evidence for a wide range of values for the cut-off energy, spanning from 30 to 300 keV and higher, suggests a possible trend of increasing cut-off energy as the power law photon index increases (Perola et al. 2002). However, it is not clear if this effect is due to limitations in the spectral analysis or if it is intrinsic to the sampled source population.

Although broad-band measurements of AGN have been made in the past, as already said mainly with the *BeppoSAX* satellite (e.g. Perola et al. 2002 and Dadina 2007), these did not generally pertain to a complete sample of sources. In this work we attempt to do this analysis for the first time on a complete sample of 36 type 1 active galaxies selected in the 20-40 keV band. First we discuss the sample selection and then present the broad-band (1-110 keV) spectral analysis of the sample sources. Next, we discuss the sample average properties and how they compare with the current framework of AGN spectral modeling and the recent synthesis of the Cosmic Diffuse X-ray Background.

## 2 THE SAMPLE

The complete sample has been extracted starting from a set of 57 type 1 AGN detected in the 20-40 keV band and listed in the $3^{rd}$ IBIS/ISGRI catalogue (Bird et al. 2007). This catalogue contains 140 objects firmly identified as AGN (63 of type 1, 64 of type 2 and 13 blazars), plus a dozen sources candidate to be active galaxies. Within the sample of type 1 AGN, 6 are narrow line Seyfert 1s and have not been considered here because of their peculiar properties when compared to their broad line analogues (Malizia et al. 2008). The remaining 57 objects are all identified as broad line active galaxies, either from NED (NASA Extragalactic Database) or from follow-up spectroscopic observations [1]. All these initial 57 AGN are Seyfert of type between 1 and 1.5, have redshifts below 0.15 and an average z of 0.05. From this list of 57 AGN, a complete sample has then been extracted by means of the $V/V_{max}$ test.

The $V/V_{max}$ test was first introduced by Schmidt (1968) as a test of uniformity of distribution in space for a flux-limited sample of objects. It can, however, be used in the opposite sense, that is, assuming that the sample is distributed uniformly in space (and that there is no evolution), it is possible to test if the sample is complete. The test consists of comparing the volumes contained within the distances where the sources are observed (V) with the maximum volumes ($V_{max}$), defined as those within the distance at which each source would be at the limit of detection. If the sample is not complete, the expected value for $<V/V_{max}>$ is less than 0.5, while when complete it should be equal to 0.5.

[1] For optical classification of *INTEGRAL* sources, including some of the AGN analysed here, please refer to Nicola Masetti's web page at http://www.iasfbo.inaf.it/extras/IGR/main.html

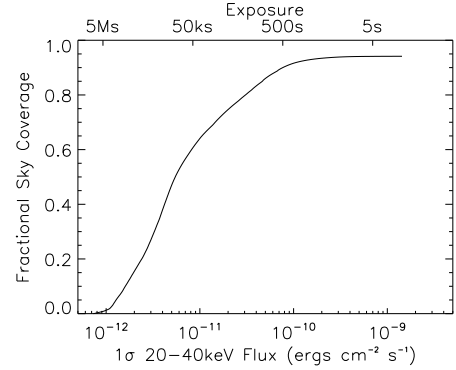

**Figure 1.** The fraction of the sky seen as a function of both $1\sigma$ limiting flux and exposure for the complete $3^{rd}$ catalogue (Bird et al. 2007). It can be seen that large fractions of the sky have very different sensitivity limits.

In the case of the ISGRI catalogue, the sky exposure, and therefore the limiting sensitivity is a strong function of position, as is shown in figure 1. This can be taken into consideration by using the $V_e/V_a$ variation of the test, introduced by Avni & Bahcall (1980). Once again the expected mean value m=$<V_e/V_a>$ will be 0.5 when the sample is complete.

For our specific case, the significance for each source given in the catalogue is not that found in the sky map, but a value which is adjusted after the source is detected and a light curve created for it (Bird et al. 2007). In applying the $V_e/V_a$ test, the significances used are those which are the basis of finding the source i.e. from the sky map.

Figure 2 shows the value of $<V_e/V_a>$ as a function of limiting sensitivity. It can be seen that the increasing trend becomes flat above about $5.5\sigma$, where the ratio is 0.45±0.05, consistent with completeness.

There are 36 type 1 AGN detected at a significance higher than this value, forming our complete sample of hard (20-40 keV) X-ray selected type 1 AGN. Their average redshift is 0.04, slightly lower than that of the entire sample of 57 objects.

In table 1, relevant parameters for each of these sources are listed (coordinates, Galactic column density, redshift values and 2-10 and 20-100 keV luminosities) together with a detailed optical classification. Objects, which are radio loud (RL) according to the definition given by Molina et al. (2008) are highlighted in bold typeface in table 1. Fluxes have been converted to luminosities, assuming $H_0$=71 km s$^{-1}$Mpc$^{-1}$ and $q_0$=0 (Spergel et al. 2003).

## 3 DATA REDUCTION

The sources presented here have all been observed by *XMM-Newton* with the exception of GRS 1734-292 and NGC 6814, observed by *ASCA*, QSO B0241+62 observed by *Chandra* and IGR J00333+6122, Swift J0917.2-6221, Mrk 50, IGR J13109-555, IGR J16119-6033, IGR J17488-3253, IGR J18259-0706, 2E 1853.7+1534 and S5 2116+81 for which *Swift*/XRT data have instead been used (see observation log in table 2). For 4U 1344-60 only MOS1 and pn data were available, while for IGR J16482-3036 only MOS1 and MOS2 data were used. In the cases of MCG+08-11-011 and IC 4329A only pn data were analysed, following Matt

**Table 1**
**Complete Sample of Type 1 AGN detected above 5.5$\sigma$ in the 20-40 keV band**

| Name | RA | DEC | $N_H$ $10^{22}$cm$^{-2}$ | $z^a$ | $L_{2-10keV}$ $10^{44}$ erg s$^{-1}$ | $L_{20-100keV}$ $10^{44}$ erg s$-1$ | Type |
|---|---|---|---|---|---|---|---|
| IGR J00333+6122 | 8.360 | 61.457 | 0.55 | 0.105 | 1.54 | 2.73 | Sy 1.5 |
| **QSO B0241+62** | **41.285** | **62.48** | **0.75** | **0.044** | **1.50** | **2.73** | **Sy 1** |
| **B3 B0309+411B** | **48.273** | **41.343** | **0.13** | **0.136** | **0.94** | **9.75** | **Sy 1/BLRG** |
| **3C 111** | **64.573** | **38.014** | **0.32** | **0.0485** | **1.78** | **4.75** | **Sy 1/BLRG** |
| LEDA 168563 | 73.028 | 49.514 | 0.54 | 0.029 | 0.82 | 1.14 | Sy 1 |
| 4U 0517+17 | 77.676 | 16.477 | 0.22 | 0.01788 | 0.15 | 4.37 | Sy 1.5 |
| MCG+08-11-011 | 88.717 | 46.442 | 0.20 | 0.0205 | 0.51 | 0.57 | Sy 1.5 |
| Mrk 6 | 103.032 | 74.423 | 0.06 | 0.018813 | 0.19 | 0.38 | Sy 1.5 |
| IGR J07597-3842 | 119.923 | -38.719 | 0.60 | 0.04 | 0.57 | 1.32 | Sy 1.2 |
| FRL 1146 | 129.620 | -36.013 | 0.40 | 0.031578 | 0.27 | 0.38 | Sy 1.5 |
| Swift J0917.2-6221 | 139.040 | -62.314 | 0.19 | 0.0573 | 1.01 | 1.52 | Sy 1 |
| NGC 3783 | 174.733 | -37.745 | 0.08 | 0.00973 | 0.12 | 0.28 | Sy 1 |
| NGC 4151 | 182.636 | 39.409 | 0.02 | 0.003319 | 0.06 | 0.15 | Sy 1.5 |
| Mrk 50 | 185.862 | 2.693 | 0.02 | 0.023433 | 0.12 | 0.17 | Sy 1 |
| NGC 4593 | 189.905 | -5.353 | 0.02 | 0.009 | 0.06 | 0.13 | Sy 1 |
| IGR J12415-5750 | 190.377 | -57.825 | 0.30 | 0.0244 | 0.13 | 0.29 | Sy 1 |
| **IGR J13109-5552** | **197.682** | **-55.863** | **0.22** | **0.0850** | **0.76** | **0.33** | **Sy 1** |
| MCG-06-30-015 | 203.995 | -34.302 | 0.04 | 0.007749 | 0.05 | 0.06 | Sy 1.2 |
| 4U 1344-60 | 206.883 | -60.610 | 1.07 | 0.013 | 0.13 | 0.27 | Sy 1.5 |
| IC 4329A | 207.339 | -30.309 | 0.04 | 0.016054 | 0.69 | 1.20 | Sy 1.2 |
| IGR J16119-6036 | 242.988 | -60.658 | 0.23 | 0.016 | 0.02 | 0.14 | Sy 1 |
| IGR J16482-3036 | 252.062 | -30.590 | 0.18 | 0.031 | 0.23 | 0.64 | Sy 1 |
| IGR J16558-5203 | 254.010 | -52.062 | 0.30 | 0.054 | 1.10 | 2.08 | Sy 1.2 |
| GRS 1734-292 | 264.377 | -29.137 | 0.77 | 0.0214 | 0.34 | 0.83 | Sy 1 |
| 2E 1739.1-1210 | 265.474 | -12.215 | 0.21 | 0.037 | 0.37 | 0.81 | Sy 1 |
| IGR J17488-3253 | 267.217 | -32.926 | 0.53 | 0.02 | 0.11 | 0.34 | Sy 1 |
| IGR J18027-1455 | 270.685 | -14.916 | 0.50 | 0.035 | 0.18 | 1.14 | Sy 1 |
| IGR J18259-0706 | 276.495 | -7.136 | 0.62 | 0.037 | 0.13 | 0.52 | Sy 1 |
| **3C 390.3** | **280.586** | **79.781** | **0.04** | **0.0561** | **1.45** | **4.06** | **Sy 1/BLRG** |
| 2E 1853.7+1534 | 283.970 | 15.618 | 0.39 | 0.084 | 1.92 | 3.64 | Sy 1 |
| NGC 6814 | 295.657 | -10.320 | 0.13 | 0.005214 | 0.001 | 0.03 | Sy 1.5 |
| **4C 74.26** | **310.585** | **75.145** | **0.12** | **0.104** | **5.40** | **10.97** | **Sy 1/BLRG** |
| **S5 2116+81** | **318.492** | **82.072** | **0.07** | **0.084** | **1.87** | **6.02** | **Sy 1/BLRG** |
| **IGR J21247+5058** | **321.172** | **50.972** | **1.11** | **0.02** | **0.45** | **0.94** | **Sy 1/BLRG** |
| MR 2251-178 | 343.543 | -17.607 | 0.03 | 0.064 | 3.62 | 6.32 | Sy 1 |
| MCG-02-58-022 | 346.200 | -8.666 | 0.04 | 0.04686 | 1.45 | 1.84 | Sy 1.5 |

*Note*: sources classified as RL are highlighted in bold typeface (see also Molina et al. 2008).
$^a$: type and redshift are taken from Nicola Masetti's webpage at http://www.iasfbo.inaf.it/extras/IGR/main.html.

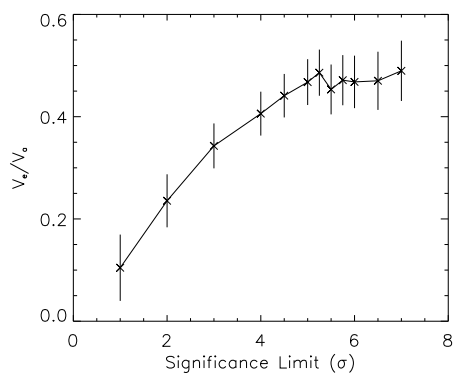

**Figure 2.** The value of $<V_e/V_a>$ as a function of limiting significance.

et al. (2006) for the former source and Markowitz et al. (2006) for the latter.

MOS and pn (Strüder et al. 2001; Turner et al. 2001) data were reprocessed using the *XMM-Newton* Standard Analysis Software (SAS) version 7.0 employing the latest available calibration files. Only patterns corresponding to single, double, triple and quadruple X-ray events for the two MOS cameras were selected (PATTERN$\leqslant$12), while for the pn only single and double events (PATTERN$\leqslant$4) were taken into account; the standard selection filter FLAG=0 was applied. Exposures have been filtered for periods of high background and the resulting exposures are listed in table 2. Source counts were extracted from circular regions of typically 40″-50″ of radius centered on the source, while background spectra were extracted from circular regions close to the source or from source-free regions of typically 20″ of radius. The ancillary response matrices (ARFs) and the detector response matrices (RMFs) were generated using the *XMM*-SAS tasks *arfgen* and *rmfgen*; spectral channels were rebinned in order to achieve a minimum of 20 counts per each bin. For sources affected by pile-up, the central 5″ of the PSF have been excised and spectra have been extracted from annular regions of typically 50″ (external radius).

In all our fits the cross calibration constants between the XMM instruments, pn/MOS1 and MOS2/MOS1, are always taken into account; these were left free to vary and always found to be in the range 0.97-1.04.

*ASCA* data (spectra and associated files) were downloaded from the TARTARUS database version 3.1[2]. The cross calibration constant between the two *ASCA*-GIS instruments has always been assumed to be 1.

*Chandra* data reduction for QSO B0241+62 was performed with CIAO 3.4 and CALDB 3.2.4 to apply the latest gain corrections. Subsequent filtering on event grade and exclusion of periods of high background resulted in a total exposure of 34 ks. The CIAO script "psextract" was used to generate the spectrum, with appropriate background and response files, of the point source; the source spectrum was extracted using a radius of $\sim 4''$ while background files were generated using a region of $\sim 9''$. With a count rate of 0.14 counts/frame, the pile-up fraction is insignificant at <10%.

XRT data reduction was performed using the standard data pipeline package (xrtpipeline v.0.10.6) in order to produce screened event files. All data were collected in the Photon Counting (PC) mode (Hill et al. 2004), adopting the standard grade filtering (0-12 for PC) according to the XRT nomenclature. Source data have been extracted using photons in a circular region of typically 20″ radius; background data have instead been taken from various uncontaminated regions near the X-ray source, using either a circular region of different radii or an annulus surrounding the source.

The *INTEGRAL* data reported here consist of several pointings performed by ISGRI (Lebrun et al. 2003) between revolutions 12 and 429, i.e. the period from launch to the end of April 2006. ISGRI images for each available pointing were generated in various energy bands using the ISDC offline scientific analysis software OSA (Goldwurm et al. 2003) version 5.1. Count rates at the position of the source were extracted from individual images in order to provide light curves in various energy bands; from these light curves, average fluxes were then extracted and combined to produce an average source spectrum (see Bird et al. 2007 for details). Analysis has been performed in the 20-110 keV band.

Table 2 reports the observation log for all 36 sources: it lists the X-ray observation date and the exposure of both X- and gamma-ray measurements.

In the present analysis we also made use of publicly available *Swift*/BAT spectra, retrieved on the web[3]; these spectra are from the first 9 months of operations of the *Swift*/BAT telescope (Baumgartner et al. 2008 and Tueller et al. 2008).

## 4  BROAD-BAND SPECTRAL ANALYSIS

The *XMM-Newton*, *Chandra*, *Swift*/XRT and *ASCA* data were fitted together with *INTEGRAL* and *Swift*/BAT points using XSPEC v.11.2.3 (Arnaud 1996); errors are quoted at 90% confidence level for one parameter of interest ($\Delta\chi^2$=2.71). Since the study of the soft excess is not an objective of the present work, but rather the understanding of the high energy emission characteristics (photon index, high energy cut-off and reflection) of our sources, the analysis has been restricted to the 1-110 keV energy range. Furthermore, in a number of the brightest objects, the fit has been done in the more restricted energy range of 3-100 keV, in order to avoid complications due to low energy features, such as warm absorber and/or soft excess, since their treatment is beyond the scope of this work. To describe the source data in this energy band we have adopted the pexrav model in XSPEC, i.e. an exponentially cut-off power law reflected from neutral material, where the reflection component is described by the parameter $R=\Omega/2\pi$, and the inclination angle has been fixed to 30°, i.e. a nearly face-on geometry as expected for type 1 sources. Galactic and intrinsic absorption are as well added to the model. A gaussian component has also been included to take into account the presence of the iron K$\alpha$ line at 6.4 keV. This addition, however, does not apply to those sources observed by the *XRT* telescope, since this instrument is not sensitive enough for the detection of the iron line feature. Whenever the width of the line was compatible with being narrow, we fixed it to 10 eV.

---
[2]  http://tartarus.gsfc.nasa.gov
[3]  http://swift.gsfc.nasa.gov/docs/swift/results/bs9mon/

We point out that, for some of the sources in the sample, the adopted model does not give a good representation of the data. There are, in fact, 5 sources (see table 3) which require more complex absorption in addition to (NGC 4151 and 4U 1344-60) or in substitution of (Mrk 6, IGR J16558-5203 and IGR J21247+5058) the neutral column density fully covering the AGN nucleus: in all these cases, the complexity is introduced by means of one or two layers of absorption partially covering the source. Finally, in the case of MCG-06-30-015, a single gaussian component is not sufficient to model the iron line which, in this source, is known to be relativistically broadened (Guainazzi et al. 1999; Fabian et al. 2002; Vaughan & Fabian 2004). To take this evidence into account, a simple parameterisation of this feature has been adopted in the form of a narrow Gaussian line, with energy and width fixed to 6.4 keV and 10 eV respectively, plus a laor model in XSPEC to represent the relativistic iron line.

In the fitting procedure, a multiplicative constant, $C$, has been introduced to take into account possible cross-calibration mismatches between the X-ray and the soft gamma-ray data (see table 3). When treating the *INTEGRAL* data, this constant $C_{IBIS}$ has been found to be close to 1 with respect to *XMM-Newton*, *Swift*/XRT and *Chandra* using various source typologies (e.g. Landi et al. 2007; Masetti et al. 2007; De Rosa et al. 2002; Panessa et al. 2008), so that significant deviation from this value can be confidently ascribed to source flux variability; also when considering BAT data, a cross-calibration constant $C_{BAT}$ different than 1 often means flux variation in the source analysed (see for example Ajello et al. 2008 and Winter et al. 2008). In figure 3, a plot of the cross-calibration constants between X-ray and BAT ($C_{BAT}$) and X-ray and ISGRI ($C_{IBIS}$) data is shown, together with the 1 to 1 line between these two quantities. Since some variances are expected between $C_{BAT}$ and $C_{IBIS}$, due to differences in observation times and average fluxes obtained/measured by BAT and ISGRI, the good match between the two cross-calibration constants is surprising and suggests that flux variability is rare at high energies in type 1 AGN (Beckmann et al. 2007). Of the 11 sources in common between their sample and ours, only NGC 4151 is reported as a variable source above 10 keV, in agreement with our finding of significantly different BAT/IBIS calibration constants. Giving a closer look, we find however some discrepancies due to the fact that, while $C_{IBIS}$ clusters around 1 (a sign that the spectral match between X-ray and ISGRI data is quite good), $C_{BAT}$ clusters at a lower value of 0.6-0.7; this is an indication that BAT spectra may have some systematics when compared to lower energy data. Only NGC 6814 has an unexpected large value of $C$ for both instruments (hence is not plotted in the figure) suggestive of a large change in flux between the X-ray and gamma-ray measurements (see also Molina et al. 2006).

The relevant parameters obtained using the adopted model are listed in table 3, together with the goodness of each fit; radio loud objects have already been discussed in detail by Molina et al. (2008), but are reconsidered here as part of the complete sample. Figures 4, 5, 6 and 7 show a set of spectral fits indicative of the large range of parameters sampled in this work.

**Table 2**
**Observations Log**

| Name | Obs. Date | Exp. (pn) (ksec) | Exp. (MOS) (ksec) | Exp. (XRT[‡]) (ksec) | Exp. (ASCA) (ksec) | Exp. (Chandra) (ksec) | Exp. (INTEGRAL[†]) (ksec) |
|---|---|---|---|---|---|---|---|
| IGR J00333+6122 | 14/11/2007 | - | - | 6.04 | - | - | 1787 |
| QSO B0241+62 | 06/04/2001 07/04/2001 | - | - | - | - | 34 | 372 |
| B3 0309+411 | 04/09/2005 | 11.0 | 19.7/16.1 | - | - | - | 5067 |
| 3C 111 | 14/03/2001 | 14.5 | 7.6/7.1 | - | - | - | 2006 |
| LEDA 168563 | 26/02/2007 | 9.6 | 10.8/9.4 | - | - | - | 157 |
| 4U 0517+17 | 22/08/2007 | 35.7 | 45.6/45.4 | - | - | - | 233 |
| MCG+08-11-011 | 09/04/2004 | 10.3 | 12.4/16.9 | - | - | - | 48 |
| Mrk 6 | 27/03/1001 | 9.0 | 14.9/7.4 | - | - | - | 581 |
| IGR J07597-3842 | 08/04/2006 | 11.9 | 14.3/13.3 | - | - | - | 1049 |
| FRL 1146 | 12/12/2006 | 6.9 | 6.3/7.1 | - | - | - | 1172 |
| Swift J0917.2-6221 | 11/11/2005 13/12/2005 | - | - | 12.9 | - | - | 1140 |
| NGC 3783 | 28/12/2000 | 32.2 | 35.0/33.9 | - | - | - | 39 |
| NGC 4151 | 27/05/2003 | 10.4 | 15.8/15.8 | - | - | - | 449 |
| Mrk 50 | 14/11/2007 15/11/2007 16/11/2007 | - | - | 18.3 | - | - | 1164 |
| NGC 4593 | 02/07/2000 | 3.5 | 6.0 (MOS2) | - | - | - | 1040 |
| IGR J12415-5750 | 01/02/2006 | 4.0 | 8.3/6.6 | - | - | - | 1604 |
| IGR J13109-5552 | 25/12/2006 28/12/2006 | - | - | 6.58 | - | - | 1288 |
| MCG-06-30-015 | 04/08/2001 | 98 | 83.7/92.1 | - | - | - | 473 |
| 4U 1344-60 | 25/08/2001 | 28.7 | 38.7 (MOS1) | - | - | - | 1510 |
| IC 4329A | 06/08/2003 | 47.2 | 44.7/41.9 | - | - | - | 382 |
| IGR J16119-6036 | 15/06/2007 | - | - | 2.5 | - | - | 1786 |
| IGR J16482-3036 | 01/03/2006 | - | 7.0/7.0 | - | - | - | 2460 |
| IGR J16558-5203 | 01/0372006 | 4.8 | 5.8/5.4 | - | - | - | 2502 |
| GRS 1734-292 | 08/09/1998 | - | - | - | 6.1 | - | 6087 |
| 2E 1739.1-1210 | 04/04/2006 | 13.1 | 14.7/14.7 | - | - | - | 1515 |
| IGR J17488-3253 | 02/11/2005 | - | - | 16.8 | - | - | 5766 |
| IGR J18027-1455 | 25/03/2006 | 14.2 | 17.5/16.9 | - | - | - | 2500 |
| IGR J18259-0706 | 12/02/2007 | - | - | 4.2 | - | - | 157 |
| 3C 390.3 | 08/10/2004 | 38.8 | 32.4/34.60 | - | - | - | 3058 |
| 2E 1853.7+1534 | 20/07/2007 29/07/2007 | - | - | 2.6 | - | - | 785 |
| NGC 6814 | 04/05/1993 | - | - | - | 41.7 | - | 1853 |
| 4C 74.26 | 06/02/2004 | 25.6 | 29.0/29.0 | - | - | - | 1213 |
| S5 2116+81 | 25/06/2006 17/10/2006 | - | - | 10.2 | - | - | 1985 |
| IGR J21247+5058 | 06/11/2005 | 22.2 | 24.2/23.6 | - | - | - | 1067 |
| MR 2251-178 | 18/05/2002 | 8.5 | 13.6/10.9 | - | - | - | 307 |
| MCG-02-58-022 | 01/12/2000 | 7.2 | - | - | - | - | 320 |

[†]: note that for INTEGRAL the exposure refer to a number of pointings in the period between launch and April 2006.
[‡]: total exposure of the summed pha files.

## 5 THE COMPLETE SAMPLE: GENERAL PROPERTIES

In the following, we discuss the general properties of our sample, using parameter distributions and mean values. We use the arithmetic mean and adopt its standard deviation as a measure of each parameter spread[4]. This is due to the extreme non-gaussian distribution of the data points and asymmetry in the errors, which are therefore not taken into account with this method. We can use these errors, however, by adopting an alternative method to estimate the averages and their uncertainties for the various parameters. We have distributed the probability of every individual point according to the upper and lower error, assuming gaussianity and 50% probability that the true value is either above or below the measured one. We then take the median of the probability distribution as the most likely value and the $1\sigma$ error (above and below the value) as those points containing 68% of the integrated probability. The results of both methods, summarised in table 4, are fully compatible within the relative uncertainties.

[4] Limits on the parameters have not been considered in this evaluation.

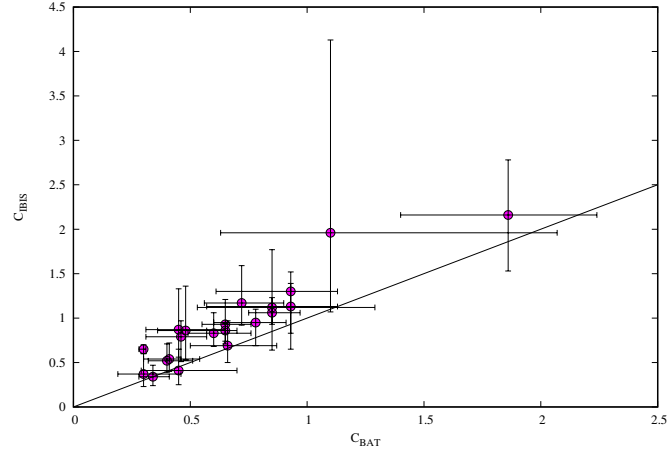

**Figure 3.** Plot of the cross-calibration constants between X and Gamma-ray data; $C_{BAT}$ is the *XMM/ASCA/XRT* and BAT cross-calibration constant and $C_{IBIS}$ is the *XMM/ASCA/XRT* and *INTEGRAL* one. The 1 to 1 ($C_{BAT}=C_{IBIS}$) line for the constants is also shown. NGC 6814 has been excluded from the plot (see text for details).

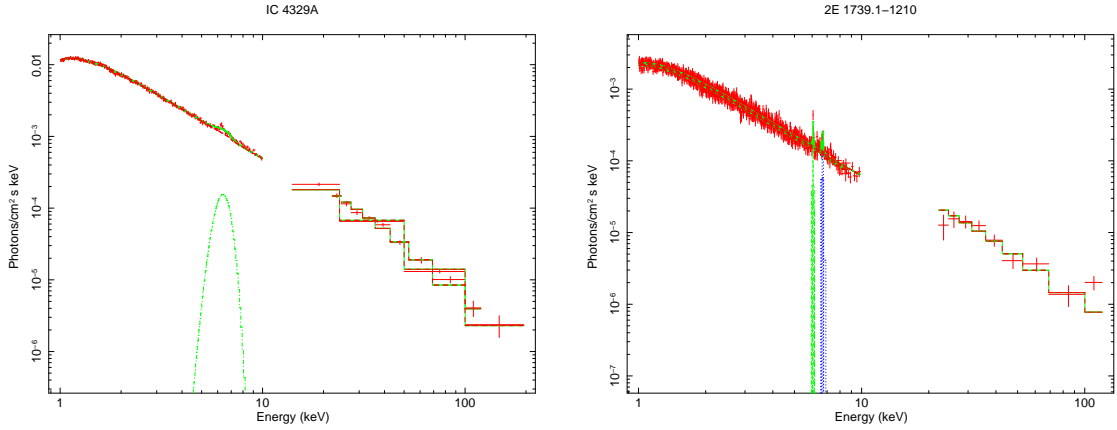

**Figure 4.** Examples of broad band spectra fitted with the `pexrav` model: IC 4329A (left panel) and 2E 1739.1-1210 (right panel). IC 4329A is characterised by the presence of a broad Fe line besides a cut-off and some reflection, while 2E 1739.1-1210 has no constraints on the cut-off energy but shows a high reflection fraction.

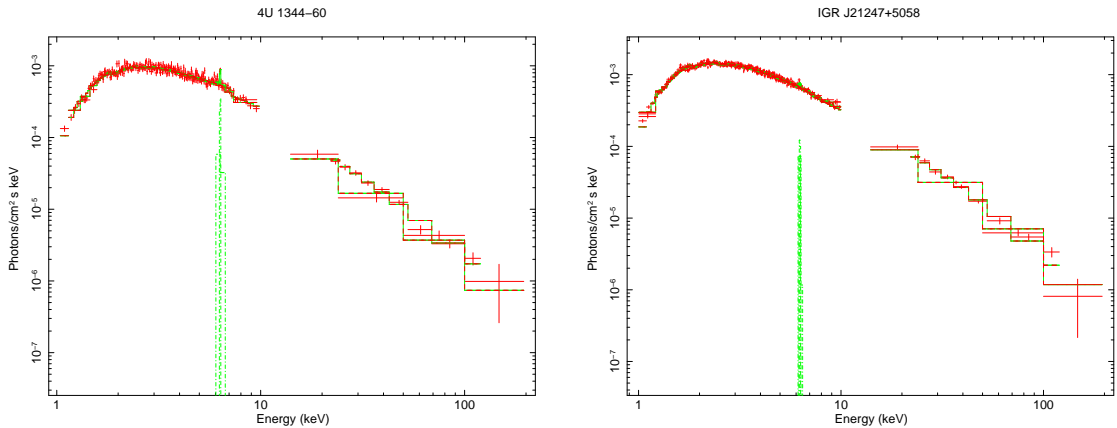

**Figure 5.** Examples of broad band spectra fitted with the `pexrav` model: 4U1344-60 (left panel) and IGR J21247+5058 (right panel). Both sources require complex absorption in the form of three layers (one totally and two partially covering the central source) in the case of 4U 1344-60 and of two layers (both partially covering the central nucleus) in the case of IGR J21247+5058.

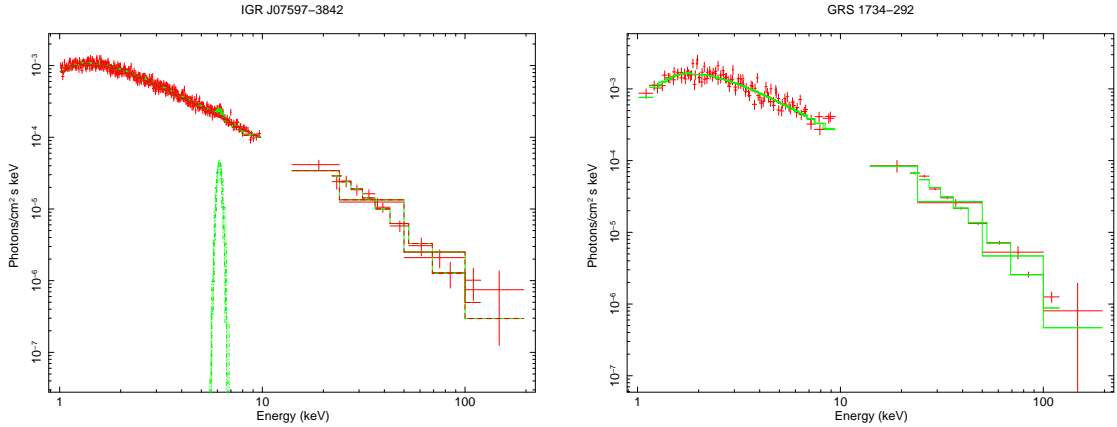

**Figure 6.** Examples of broad band spectra fitted with the `pexrav` model: IGR J07597-3842 (left panel) and GRS 1734-292 (right panel). Both sources are characterised by a very low cut-off energy, around 70 keV for IGR J07597-3842 and around 40 for GRS 1734-292.

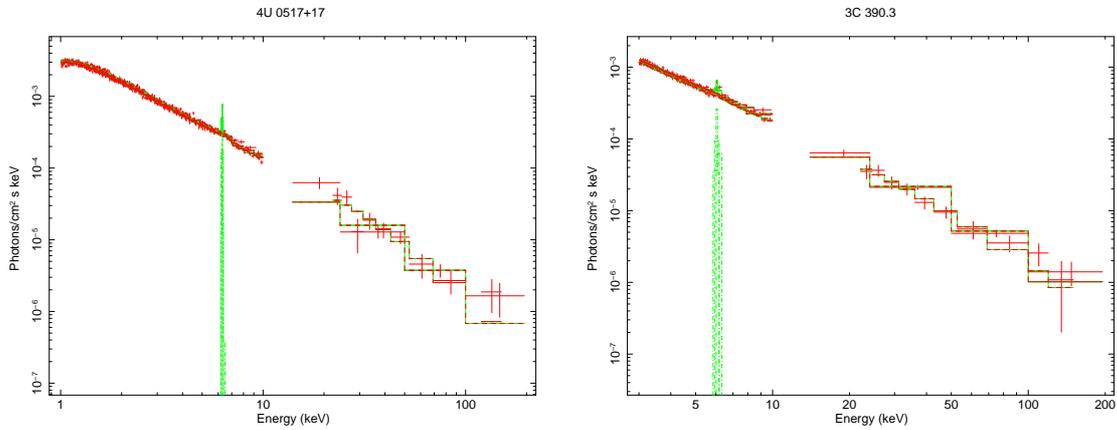

**Figure 7.** Examples of broad band spectra fitted with the `pexrav` model: 4U 0517+17 (left panel) and 3C 390.3 (right panel). Both sources are characterised by cut-off energies above 100 keV; 4U 0517+17 has also a high value of R (above 2).

Almost 60% of the sample sources, despite being type 1 AGN, require absorption in excess of the Galactic value; for these absorbed objects, the majority (86%) have a simple absorber fully covering the source, while 24% require (in addition or in substitution) a more complex type of absorption. The distribution of the intrinsic column densities is shown in figure 8, where for clarity only measured $N_H$ values, as reported in table 3, have been considered. Note that, for objects having complex absorption, the value of the intrinsic column density is relative to the layer with the highest $N_H$. The absorption distribution shows a peak around $Log(N_H)$=21-22[5], with the sources characterised by complex absorption occupying the tail towards higher values of $N_H$. Also Winter et al. (2009) find that complex absorption is associated to sources with the highest column densities, but their $N_H$ values are lower than ours. We have 5 sources showing complex absorption (NGC 4151, Markarian 6, 4U 1344-60, IGR J16558-5203 and IGR J21247+5058) and all of them have been individually discussed in the literature (De Rosa et al. 2007, Malizia et al. 2003, Piconcelli et al. 2005, Panessa et al. 2008, Molina et al. 2007); our results are in full agreement with these previous studies and further confirm the peculiarities of these sources. The presence of strong absorption up to 2-5 $\times 10^{23}$ cm$^{-2}$ in these AGN is at odds with their type 1 classification, but this discrepancy can be circumvented if the "clumpy torus" model, recently proposed by Elitzur & Schlosman (2006), is applied to our 5 galaxies. In this model, the torus is not a continuous toroidal structure, but is rather made of clouds with $N_H$ in the range $10^{22} - 10^{23}$ cm$^{-2}$, distributed around the equatorial plane of the AGN. The broad line region (BLR) represents the inner segment of the torus and this is the region where the X-ray absorption is likely to occur. Because of this clumpiness, the difference between type 1 and 2 AGN is not only due to orientation but also to the probability of direct viewing of the active nucleus or, in other words, on how many clouds our line of sight intercepts. The classification as intermediate Seyferts for 4 out of 5 objects[6] suggests that in these galaxies we might be viewing the nucleus at an intermediate angle, so that a greater fraction of the BLR clouds is intercepted and a higher absorption measured; variations in the absorption properties as ob-

---

[5] If Galactic $N_H$ values were to be taken into consideration as upper limits to the absorption, the peak at $Log(N_H)$=21-21.5 would be more evident and the average column density values lower.

[6] The first 3 AGN are of type 1.5, IGR J16558-5203 is of type 1.2, while the classification of IGR J21247+5058 is less well defined, since the only broad optical line visible is superimposed on a Galactic star continuum (Molina et al. 2007)

**Table 3**

**AGN Complete Sample-pexrav model**

| Name | $N_H^{FC}$ ($10^{22}$ cm$^{-2}$) | $N_H^{PC}$ (cf) ($10^{22}$ cm$^{-2}$) | $\Gamma$ | $E_c$ (keV) | R | EW† (eV) | $\sigma$ (eV) | $C_{BAT}$ | $C_{IBIS}$ | $\chi^2$ (dof) |
|---|---|---|---|---|---|---|---|---|---|---|
| IGR J00333+6122 | $0.85^{+0.45}_{-0.40}$ | - | $1.76^{+0.30}_{-0.28}$ | NC | <1.15 | - | - | - | $0.95^{+0.97}_{-0.60}$ | 18.4 (28) |
| **QSO B0241+61** | $0.21^{+0.15}_{-0.10}$ | - | $1.64^{+0.22}_{-0.14}$ | >86 | $0.56^{+0.82}_{-0.08}$ | $73^{+44}_{-42}$ | 10f | $0.85^{+0.44}_{-0.32}$ | $1.12^{+0.65}_{-0.48}$ | 222.2 (233) |
| **B3 0309+411** | - | - | $1.90^{+0.08}_{-0.08}$ | $35^{+91}_{-17}$ | $3.48^{+1.58}_{-2.24}$ | $59^{+42}_{-43}$ | 10f | - | $6.89^{+3.90}_{-3.38}$ | 554.2 (601) |
| **3C 111** | $0.46^{+0.03}_{-0.03}$ | - | $1.73^{+0.06}_{-0.06}$ | $126^{+193}_{-50}$ | $0.85^{+0.57}_{-0.08}$ | <30 | 10f | $0.40^{+0.11}_{-0.08}$ | $0.52^{+0.19}_{-0.13}$ | 1227.9 (1497) |
| LEDA 168563★ | - | - | $1.69^{+0.12}_{-0.11}$ | $114^{+511}_{-60}$ | <1.24 | $24^{+18}_{-15}$ | 10f | $0.45^{+0.20}_{-0.17}$ | $0.87^{+0.46}_{-0.31}$ | 967.3 (1298) |
| 4U 0517+17 | $0.09^{+0.02}_{-0.01}$ | - | $1.83^{+0.04}_{-0.03}$ | >118 | $2.74^{+0.51}_{-0.41}$ | $66^{+13}_{-7}$ | 10f | $0.41^{+0.13}_{-0.11}$ | $0.54^{+0.18}_{-0.14}$ | 2519.6 (2436) |
| MCG+08-11-011 | - | - | $1.86^{+0.06}_{-0.02}$ | >126 | $2.54^{+0.53}_{-0.54}$ | $92^{+15}_{-29}$ | 10f | $0.34^{+0.13}_{-0.11}$ | $0.34^{+0.13}_{-0.10}$ | 1012.2 (1263) |
| Mrk 6‡ | - | $2.04^{+0.61}_{-0.49}$ ($0.91^{+0.02}_{-0.02}$) $8.12^{+4.76}_{-2.83}$ ($0.49^{+0.10}_{-0.06}$) | $1.42^{+0.20}_{-0.14}$ | $82^{+202}_{-43}$ | <1.22 | $57^{+23}_{-23}$ | 10f | $0.93^{+0.06}_{-0.36}$ | $1.13^{+0.26}_{-0.48}$ | 560.7 (550) |
| IGR J07597-3842 | - | - | $1.56^{+0.03}_{-0.03}$ | $70^{+30}_{-13}$ | $0.92^{+0.67}_{-0.30}$ | $89^{+31}_{-35}$ | $192^{+99}_{-79}$ | $0.78^{+0.18}_{-0.18}$ | $0.95^{+0.15}_{-0.26}$ | 435.7 (458) |
| FRL 1146 | $0.28^{+0.07}_{-0.08}$ | - | $1.71^{+0.13}_{-0.13}$ | $59^{+194}_{-31}$ | $1.31^{+0.45}_{-1.27}$ | $127^{+45}_{-52}$ | $110^{+109}_{-69}$ | - | $0.71^{+0.57}_{-0.30}$ | 498.0 (576) |
| Swift J0917.2-6221 | $0.41^{+0.13}_{-0.14}$ | - | $1.60^{+0.14}_{-0.12}$ | $83^{+210}_{-45}$ | <10.65 | - | - | $0.30^{+0.16}_{-0.14}$ | $0.37^{+0.23}_{-0.14}$ | 135.3 (122) |
| Mrk 50 | <0.12 | - | $1.94^{+0.06}_{-0.12}$ | >25 | <1.76 | - | - | - | $1.23^{+0.57}_{-0.41}$ | 159.6 (157) |
| NGC 3783★ | $1.16^{+0.37}_{-0.37}$ | - | $1.75^{+0.05}_{-0.09}$ | $98^{+79}_{-34}$ | $1.23^{+0.63}_{-0.57}$ | $87^{+11}_{-9}$ | 10f | $0.65^{+0.11}_{-0.10}$ | $0.93^{+0.28}_{-0.22}$ | 1605.3 (1768) |
| NGC 4151★ | $5.28^{+0.74}_{-1.22}$ | $22.44^{+13.67}_{-10.42}$ ($0.25^{+0.21}_{-0.07}$) | $1.77^{+0.06}_{-0.05}$ | $307^{+245}_{-94}$ | <0.20 | $56^{+5}_{-7}$ | 10f | $0.30^{+0.01}_{-0.02}$ | $0.65^{+0.05}_{-0.05}$ | 2037.8 (1989) |
| NGC 4593 | - | - | $1.92^{+0.01}_{-0.01}$ | >222 | $1.77^{+0.85}_{-0.76}$ | $91^{+37}_{-37}$ | 10f | $0.60^{+0.16}_{-0.11}$ | $0.83^{+0.23}_{-0.15}$ | 732.4 (855) |
| IGR J12415-5750 | - | - | $1.62^{+0.05}_{-0.05}$ | >67 | $1.85^{+1.19}_{-1.01}$ | $81^{+42}_{-59}$ | 10f | $0.45^{+0.23}_{-0.16}$ | $0.41^{+0.20}_{-0.16}$ | 576.4 (681) |
| **IGR J13109-5552** | <0.46 | - | $1.55^{+0.31}_{-0.26}$ | >58 | <5.1 | - | - | - | $1.83^{+3.56}_{-1.03}$ | 45.5 (53) |
| MCG-06-30-015★ | $1.48^{+0.33}_{-0.37}$ | - | $2.18^{+0.10}_{-0.11}$ | >76 | $0.78^{+0.52}_{-0.52}$ | $21^{+7‡}_{-8}$ | 10f | $0.72^{+0.18}_{-0.16}$ | $1.17^{+0.42}_{-0.25}$ | 1296.1 (1327) |
| 4U 1344-60 | $1.02^{+0.16}_{-0.25}$ | $5.17^{+2.28}_{-2.36}$ ($0.50^{+0.09}_{-0.11}$) $46.76^{+32.37}_{-21.08}$ ($0.40^{+0.12}_{-0.14}$) | $1.81^{+0.20}_{-0.21}$ | >108 | <0.80 | $57^{+23}_{-21}$ | 10f | $0.46^{+0.11}_{-0.15}$ | $0.79^{+0.18}_{-0.28}$ | 691.2 (764) |
| IC 4329A | $0.35^{+0.01}_{-0.01}$ | - | $1.81^{+0.03}_{-0.03}$ | $152^{+51}_{-32}$ | $0.70^{+0.35}_{-0.31}$ | $181^{+30}_{-33}$ | $443^{+84}_{-60}$ | $0.85^{+0.12}_{-0.10}$ | $1.06^{+0.17}_{-0.13}$ | 1699.4 (1617) |
| IGR J16119-6036 | - | - | $1.84^{+0.17}_{-0.24}$ | >54 | NC | - | - | - | $5.11^{+5.98}_{-2.60}$ | 17.7 (15) |
| IGR J16482-3036 | $0.10^{+0.06}_{-0.07}$ | - | $1.65^{+0.10}_{-0.15}$ | $97^{+158}_{-50}$ | $1.53^{+1.84}_{-1.51}$ | $82^{+45}_{-61}$ | 10f | - | $0.38^{+0.36}_{-0.12}$ | 439.9 (495) |
| IGR J16558-5203 | - | $18.56^{+16.17}_{-9.10}$ ($0.32^{+0.06}_{-0.07}$) | $2.20^{+0.17}_{-0.06}$ | NC | $2.20^{+1.08}_{-0.96}$ | $68^{+44}_{-43}$ | 10f | - | $1.27^{+0.26}_{-0.21}$ | 607.4 (775) |
| GRS 1734-292 | >0.21 | - | $1.36^{+0.05}_{-0.07}$ | $39^{+17}_{-7}$ | <0.84 | - | - | $0.93^{+0.20}_{-0.32}$ | $1.30^{+0.22}_{-0.47}$ | 166.1 (111) |
| 2E 1739.1-1210 | $0.15^{+0.03}_{-0.03}$ | - | $2.05^{+0.07}_{-0.06}$ | NC | $1.81^{+1.02}_{-0.84}$ | $26^{+36}_{-19}$ | 10f | - | $1.02^{+0.20}_{-0.17}$ | 815.5 (870) |
| IGR J17488-3253 | $0.34^{+0.10}_{-0.11}$ | - | $1.59^{+0.15}_{-0.15}$ | >118 | <1.14 | - | - | - | $1.63^{+0.40}_{-0.75}$ | 136.9 (136) |
| IGR J18027-1455 | $0.30^{+0.08}_{-0.08}$ | - | $1.55^{+0.13}_{-0.11}$ | $192^{+167}_{-63}$ | $0.90^{+0.60}_{-0.35}$ | $117^{+37}_{-31}$ | 10f | - | $1.72^{+0.78}_{-0.56}$ | 818.6 (898) |
| IGR J18259-0706 | $1.06^{+0.73}_{-0.61}$ | - | $1.70^{+0.42}_{-0.40}$ | NC | NC | - | - | - | $0.31^{+1.19}_{-0.19}$ | 25.4 (26) |
| **3C 390.3★** | - | - | $1.62^{+0.09}_{-0.08}$ | $152^{+274}_{-68}$ | <1.14 | $44^{+13}_{-10}$ | 10f | $0.66^{+0.21}_{-0.16}$ | $0.69^{+0.28}_{-0.19}$ | 1272.7 (1561) |
| 2E 1853.7+1534 | <0.39 | - | $1.47^{+0.41}_{-0.41}$ | $53^{+295}_{-21}$ | <11.40 | - | - | - | $1.33^{+0.90}_{-0.83}$ | 39.6 (45) |
| NGC 6814 | - | - | $1.80^{+0.09}_{-0.09}$ | $116^{+203}_{-52}$ | $7.04^{+12.43}_{-3.90}$ | $315^{+217}_{-219}$ | 10f | $7.62^{+2.87}_{-2.19}$ | $12.52^{+6.98}_{-3.41}$ | 201.2 (236) |
| **4C 74.26** | $0.14^{+0.02}_{-0.03}$ | - | $1.79^{+0.06}_{-0.07}$ | $100^{+680}_{-52}$ | $1.22^{+0.69}_{-0.70}$ | $88^{+23}_{-20}$ | $183^{+84}_{-65}$ | $0.48^{+0.18}_{-0.12}$ | $0.86^{+0.50}_{-0.33}$ | 1976.0 (2060) |
| **S5 2116+81** | <0.24 | - | $2.03^{+0.26}_{-0.20}$ | >85 | <4.0 | - | - | $1.10^{+0.97}_{-0.47}$ | $1.96^{+2.17}_{-0.89}$ | 88.2 (125) |
| **IGR J21247+5058** | - | $0.77^{+0.18}_{-0.13}$ ($0.89^{+0.10}_{-0.06}$) $7.86^{+2.02}_{-1.66}$ ($0.27^{+0.04}_{-0.05}$) | $1.48^{+0.06}_{-0.06}$ | $79^{+23}_{-15}$ | <0.21 | <30 | 10f | $0.65^{+0.11}_{-0.08}$ | $0.86^{+0.12}_{-0.12}$ | 2276.7 (2555) |
| MR 2251-178★ | $2.15^{+0.64}_{-1.21}$ | - | $1.68^{+0.10}_{-0.21}$ | $109^{+104}_{-44}$ | <0.35 | $45^{+19}_{-19}$ | 10f | $1.86^{+0.38}_{-0.46}$ | $2.16^{+0.62}_{-0.63}$ | 740.6 (862) |
| MCG-02-58-022 | - | - | $1.88^{+0.01}_{-0.02}$ | NC | $2.90^{+1.00}_{-0.77}$ | $45^{+25}_{-27}$ | 10f | - | $0.38^{+0.09}_{-0.09}$ | 856.3 (975) |

*Note*: sources classified as RL are highlighted in bold typeface.
A: $N_H^{FC}$ refers to the column density fully covering the nucleus; $N_H^{PC}$ refer to the column density partially covering the nucleus, with cf being the covering fraction
$C_{BAT}$: cross-calibration constant between *XMM/Chandra/XRT/ASCA* and BAT.
$C_{IBIS}$: cross-calibration constant between *XMM/Chandra/XRT/ASCA* and *INTEGRAL*.
★: analysed in the 3-110 keV energy range.
†: EW for the 6.4 iron K$\alpha$ line
‡: EW of the narrow line component. The EW of the broad line component is $312^{+46}_{-56}$, with $\sigma=354^{+47}_{-43}$.
The relativistic iron line is represented by the laor model in XSPEC. The other parameters of the model (fixed in fitting procedure) are:
$R_{IN}=1.23r_g$, $R_{OUT}=400r_g$, i=30°, $\beta$=-3.0

served in Markarian 6, NGC 4151 and possibly IGR J21247+5058 (De Rosa et al. 2007, Malizia et al. 2003, Molina et al. 2007) are expected in this scenario as the torus structure changes due to cloud motion.

The average value of the absorbing column density in our complete sample is around 0.3-0.5 ×10$^{22}$ cm$^{-2}$, with a larger or smaller spread depending on whether complex absorption objects are included or not; the higher mean value of $N_H$=5.4×10$^{22}$ cm$^{-2}$ obtained using the arithmetic mean (see table 4) is clearly due to the large spread found when complex absorption objects are considered. Our mean values are close to the average column density reported by Winter et al. (2009) for broad line AGN in their BAT sample; much higher average values were instead found by Dadina (2008) and Mushotzky (1984) for the broad line AGN listed in their works, possibly due to the presence of complex absorption sources such as NGC 4151 in these samples. It is however difficult to compare results when different fitting models/approaches are used: for example, both Dadina (2007, 2008) and Mushotzky (1984) prefer only one layer of absorption fully covering the source, while Win-

ter et al. (2009) use in alternative also a partial covering absorber and further restrict their analysis below 10 keV. We instead consider more than one layer (fully or partially covering the nucleus), if statistically required by the data. This has generally the effect of steepening the power law photon index with respect to less complex absorption models, so that even a comparison of photon indices in these cases is not straightforward. Besides, Dadina (2008) and Mushotzky (1984) used randomly chosen samples of AGN observed at high energies, while Winter et al. and ourselves employ sets of objects specifically selected above 10 keV. We therefore conclude that a comparison of the absorption (but also of other) properties between samples is complex and should be approached with some caution.

The distribution of the photon indices for our entire sample is shown in figure 9, where it can be seen that the majority of the sources fall in the range 1.6-1.9. Indeed, the mean photon index for the whole sample is 1.7 with a typical spread of 0.2. This mean $\Gamma$ value is slightly flatter than the canonical value of 1.9 generally assumed for type 1 AGN; it is also different from the value of

**Table 4**

**Average parameter values**

| Parameter | Standard Method | New Method |
|---|---|---|
| $N_H^\dagger$ | 5.4±11.3 | $0.46^{+8.85}_{-0.32}$ |
| $N_H^\ddagger$ | 0.6±0.6 | $0.35^{+0.89}_{-0.22}$ |
| $\Gamma$ | 1.74±0.20 | $1.76^{+0.22}_{-0.24}$ |
| $E_c$ | 109±63 | $106^{+186}_{-61}$ |
| $R^\dagger$ | 1.9±1.5 | $1.5^{+1.5}_{-1.0}$ |
| $R^\star$ | 1.5±0.7 | $1.4^{+1.3}_{-0.9}$ |
| EW | 83±61 | $125^{+160}_{-52}$ |

$\dagger$: all data (apart from limits) have been considered.

$\ddagger$: objects with complex absorption removed from the calculation

$\star$: NGC 6814 and B3 0309+411 removed from the calculation

1.89±0.03 found by Dadina (2008), who employed the same model used in this analysis but using a sample of 43 type 1 AGN seen by *BeppoSAX*. This difference may be related to the fact that our study makes use for the first time of a complete sample of hard X-ray selected AGN, whereas Dadina (2008) employed bright randomly chosen AGN. In contrast to his result, Winter et al. (2009), using 0.3-12 keV data, provide an average $\Gamma$ of ~1.7 for their sample of Seyfert 1s; the same photon index was also reported by Mushotzky (1984) for a set of broad line AGN observed over the 0.5-100 keV energy range and by Sazonov et al. (2008) for a sample of type 1 and 2 AGN in the local Universe measured over the 3-300 keV band. Even though a straightforward comparison is not possible, because of the use of different models by the various authors, the resulting similarity in $\Gamma$ is reassuring, particularly for samples selected in the same way as ours, i.e. above 10 keV.

It is also possible that the better throughput of the *XMM* instruments, with respect to the *BeppoSAX*/MECS detectors employed for the Dadina sample, provides a better description of the source spectral shape and hence a flatter spectrum. Indeed, Bianchi et al. (2009), using a catalogue of Seyferts in the *XMM/Newton* archive obtained a similar mean value (1.73) but a larger spread, even taking into account a reflection component in their spectral fits.

The mean value for the high energy cut-off is close to 110 keV with a small spread; the distribution of values for this parameter shows that most objects fall within the range 50 to 150 keV with only two AGN displaying higher cut-off energies (see figure 10). The value found in the present analysis is lower than the mean value of 230 keV found by Dadina (2008) and also than the values (around 200 keV) provided by the mean *BeppoSAX* spectrum obtained by Malizia et al. (2003). Again this maybe related to the different sample used here and the better quality of the X-ray data available for most of our objects. Despite the large spread found, we have nevertheless been able to restrict this parameter space considerably with respect for example to the Dadina's study.

The sample has also been tested against the correlation found by a number of authors between the photon index and the high energy cut-off. First claimed by Piro (1999), it was further discussed by Matt (2001) and Petrucci et al. (2001), taking advantage of the broad-band *BeppoSAX* observations of AGN. At present, this correlation is still debated, since previous works found that the two parameters are not independent in the fitting procedure, with the cut-off energy increasing as the power law steepens (Perola et al. 2002).

Figure 11 shows a plot of the photon index versus the high energy cut-off, but no evident trend is found between these two quantities. The use of a Pearson statistical test to the two sets of data also returns a low correlation coefficient of $r\sim0.12$[7] (if upper limits are ignored $r$ is 0.28); note that this test does not take into account the relative uncertainties on the values of the parameters. Studying the distribution in the parameter space of these two quantities is also an indirect way of testing that the results of our analysis are not strongly affected by the interplay between $\Gamma$ and $E_c$.

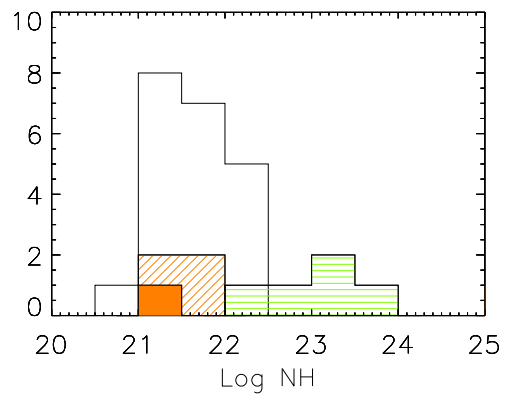

**Figure 8.** Intrinsic Column Density Distribution for the sample analysed here. The horizontally hatched histogram represents sources requiring complex absorption, for which the value of $N_H$ with the larger covering fraction has been used. The diagonally hatched histogram represents sources for which only upper limits on the parameter are available. The filled histogram is relative to GRS 1734-292, for which only a lower limit on the $N_H$ value is available.

As far as the reflection fraction is concerned, the mean value found in this study is in the range 1.5-1.9, higher than the value of 1.23 reported by Dadina (2008); our distribution of the reflection fractions shows a peak at around 1 (figure 12), with a tail extending to values of $R$ as high as 11.5. Indeed, some sources in our sample show a reflection component $R$ much greater than 1. These high values of $R$ have been found in previous studies, using *ASCA* and *BeppoSAX* data (Cappi et al. 1996; Dadina 2008) and, more recently, with the *Suzaku* satellite (Miniutti et al. 2007; Comastri et al. 2007). Such strong reflection might be present when more primary X-ray radiation is emitted toward the reflector than toward the observer, as is possible in the case of strongly variable nuclear emission or when there is a time delay between the underlying continuum and the reflected component, caused by a large distance between the reflecting material and the primary source (Malzac & Petrucci 2002; Panessa et al. 2008). Another explanation might be a peculiar geometry (Malzac et al. 2001; Malzac 2001) or general relativistic light bending effects (Fabian et al. 2004; Miniutti & Fabian 2004; Fabian et al. 2005). Recently, Gandhi et al. (2007) have shown that, in the synthesis models of the X-ray background spectrum, a significant fraction of this type of source is needed when light bending effects are taken into account. A few sources in our sample, like NGC 6814 and probably B3 0309+411, have high values of the reflection fraction simply because there is a large mismatch between X-ray and soft gamma ray data due to flux variability (Molina et al. 2006, Molina et al. 2008). Others like Swift J0917.2-6221, IGR J13109-5552, 2E 1853.7+1534 and S5 2116+81, have only loose upper limits on $R$, mostly due to the poor quality of the X-ray data. If NGC 6814 and B3 0309+411 are removed from the list of objects, the average values of $R$ become closer to the mean obtained by Dadina (2008) and the spread measured is smaller.

---

[7] The square of the correlation coefficient $r$ is conventionally used as a measure of the association between two variables.

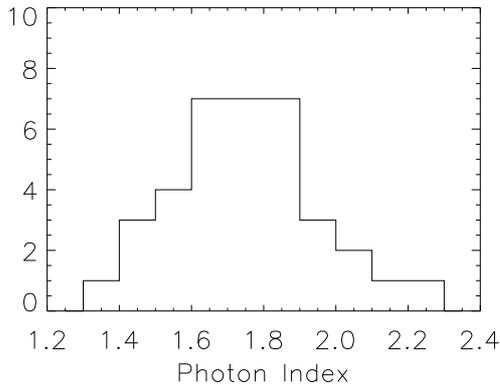

**Figure 9.** Photon index distribution of the type 1 sources analysed here.

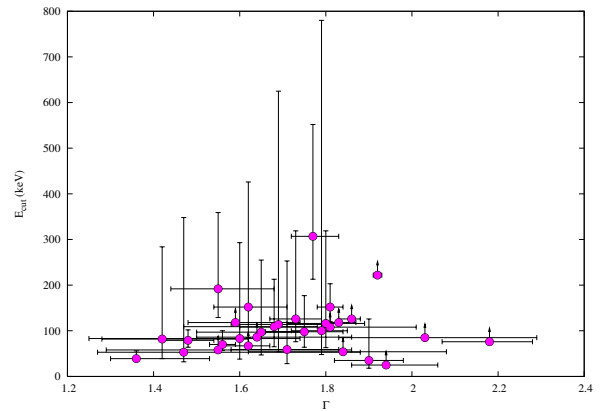

**Figure 11.** High energy cut-off vs. photon index.

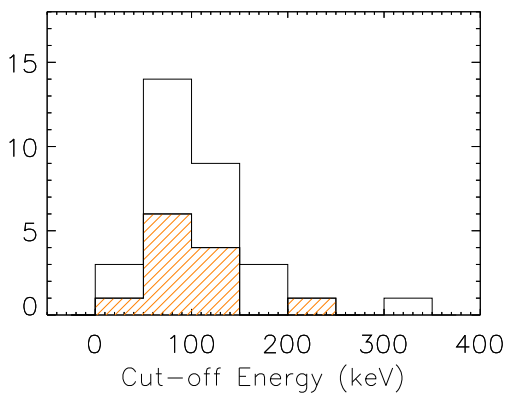

**Figure 10.** High energy cut-off distribution for the type 1 sources presented in this work. The diagonally hatched histogram represents sources for which only lower limits on E$_c$ are available.

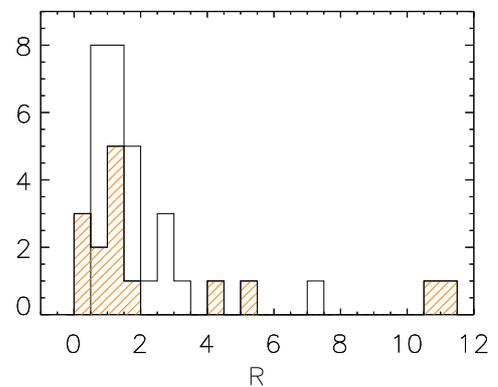

**Figure 12.** Reflection fraction distribution for all objects in the sample. The diagonally hatched histogram represents sources for which only upper limits on $R$ are available.

Since $R$ and the cross-calibration constant are strongly related in the fitting procedure, it is also important to test if a correlation exists between these two quantities. Indeed, from figure 13 (left panel), where $R$ is plotted against $C_{IBIS}$, it can be inferred that a correlation between the reflection fraction and the cross-calibration constant exists, as also suggested by applying a Pearson test to the parameter space of figure 13, which returns a strong correlation coefficient of $r$∼0.84 (not considering upper limits). However if NGC 6814 and B3 0309+411 (and all the upper limits) are removed (see figure 13 right panel) $r$ becomes ∼-0.54, implying that these two sources are those providing a high value of $r$. Furthermore, if all upper limits are taken into account, then $r$∼0.32, with NGC 6814 and B3 0309+411 included, and ∼-0.01 with them removed. This suggests that the estimate on $R$ is not very dependent on the value of $C_{IBIS}$ and that at most a very weak correlation is present between these two quantities.

A correlation between the photon index and the reflection fraction was also investigated. Zdziarski et al. (1990) proposed that such a correlation might be explained assuming that the cold medium, responsible for the reflection and surrounding the corona, affects the hardness (i.e. the slope) of the X-ray spectrum. In this framework, the soft photons, emitted from the cold medium and irradiating the X-ray source, serve as seeds for Compton upscattering. This implies that the larger the solid angle subtended by the reflector, the stronger is the flux of the soft photons and therefore the stronger the cooling of the plasma. In the case of thermal plasma, the larger the cooling by seed photons, the softer the resulting X-ray power law spectrum. A correlation between Γ and *R* has also been seen by Petrucci et al. (2001), who, employing a `pexrav` model, confirmed that the higher the temperature of the corona (which is directly linked also to the cut-off energy), the larger the reflection fraction. However, these authors, found also that using a different model that takes into account anisotropic Comptonisation, the trend is reversed and larger reflection corresponds to lower temperatures. Besides, the two parameters *R* and Γ are strongly linked in the fitting procedure and therefore a trend in the parameter space may not be entirely physical. Figure 14 is a plot of *R* vs. Γ for our set of AGN; the Pearson test applied to the two data sets gives in our case a small correlation coefficient of ~0.18, which becomes even smaller (and negative) if upper limits are considered (*r*=-0.10). This is contrary to what found by Dadina (2008), who found a correlation between Γ and *R*.

We have also looked for a correlation between the reflection fraction and the high energy cut-off (see figure 15) by applying again the Pearson test to our data: the correlation coefficient is ~-0.18 (without considering the upper/lower limits on R and $E_c$) and -0.26 when upper/lower limits are instead account for, again indicative of no correlation between these spectral parameters.

We can therefore conclude that we do not find any evidence for correlations Γ, *R* and $E_c$, an indication that the interplay between these quantities does not have an important part in the fitting procedure and has not introduced spurious relationships in our analysis.

We have also searched for a correlation between Γ, R and $E_c$ with source luminosity (considering both 2-10 keV and 20-100 keV data) but found none.

The distribution of the iron K$\alpha$ line equivalent widths is shown in figure 16. The mean value of EW is in the range 70-80 eV, with a typical spread of around 50-60 eV; this is fully compatible with the average EW (76 eV) observed by Bianchi et al. (2009) in their sample of *XMM*-Newton observed AGN. Our value is instead lower than the average of 177 keV found by Winter et al. (2009) using only the type 1 AGN in their sample; their spread is however large (142 eV) so that our result overlaps theirs. We also found that most sources have narrow Fe K$\alpha$ lines, with only ~15% of the sample requiring a broad line. This fraction is lower than the one found by Guainazzi et al. (2006) for a sample of AGN observed by *XMM*, in which at least 25% of the objects require a broad iron line. This difference maybe due to the lack of good quality X-ray data for some of our sources, i.e. those with only XRT observations available, or to the fact that, while our sample is complete, that of Guainazzi et al. (2006) is a collection of sources which happen to be observed by *XMM*. We also searched the data for the existence of an anti-correlation between EW and source luminosity, the so-called "X-ray Baldwin" or "Iwasawa-Taniguchi" effect (see Malzac & Petrucci 2002 and references therein); we could not find any evidence for this effect using our measured EW values and either 2-10 keV or 20-100 keV luminosities. This maybe due to the smaller range of luminosities probed here compared to the sample of Bianchi et al. (2007); in any case our result agrees with no evidence for the inverse Baldwin effect also reported by Winter et al. (2008).

If the iron line emission is entirely associated with the optically thick material of the disk, one would expect the line EW to correlate with the reflection fraction. Figure 17 shows a plot of *R* vs. EW for the entire sample, with radio loud objects separated from radio quiet sources; the Pearson test applied to these two sets of data, excluding upper limits, returns a correlation coefficient of only ~0.58 (if upper limits are considered *r* is 0.64), suggesting, if

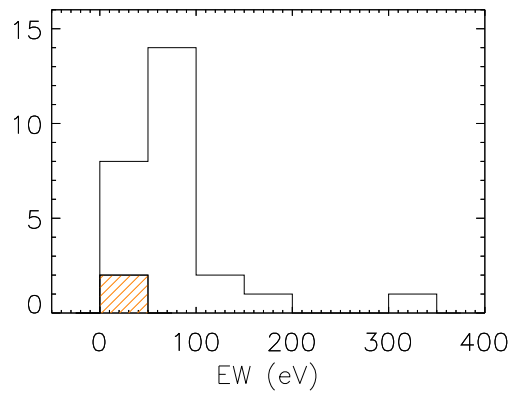

**Figure 16.** Iron line Equivalent Width (EW) distribution for all AGN in the sample. The diagonally hatched histogram represents sources for which only upper limits on EW are available.

anything, a weak correlation. In some cases like B3 0309+411, 4C 74.26, MCG+08-11-011, 2E 1739.1-1210 and MCG-02-58-022, the observed iron line EW is too small for the reflection measured: a possible explanation resides in the scatter expected in the EW values due to a variation in the iron abundance, or to a possible anisotropy of the source seed photons which might affect the observed spectrum (Petrucci et al. 2001; Merloni et al. 2006). For a given value of *R*, the EW could also differ according to the value of the power law photon index: up to Γ=2, the EW decreases as the spectrum steepens, while above Γ=2 the trend is reversed (Perola et al. 2002; Mattson, Weaver and & Reynolds 2007; Panessa et al. 2008). B3 0309+411, 4C 74.26, MCG+08-11-011, 2E 1739.1-1210 and MCG-02-58-022 have all steep spectra and so an EW lower than expected on the basis of the measured *R* value is not surprising. As shown by the results reported here, and as already observed by a number of authors (for instance Grandi et al. 2006 and Sambruna et al. 1999), the reprocessing features in BLRG tend to be, on average, quite weak with respect to their radio quiet counterparts. In the EW vs. reflection plot shown in figure 17, it is evident that radio loud AGN are more confined to a region of the plot characterised by low values of EW and to a lesser extent of *R*, whereas radio quiet objects tend to be more spread over both axes.

## 6 THE COMPLETE SAMPLE: SPECTRAL MODELING

Determining the photon indices and cut-off energies of a large, complete sample of AGN is very important for spectral modeling. So far, models have generally focused on how to reproduce and explain the observed primary continuum. A good fraction of the proposed models ascribe the power law to the inverse Compton scattering of soft photons off hot electrons. Variations to this baseline model depend on the energy distribution of these electrons and their location in relation to the accretion disk, often a hot corona above the disk. Within this scenario, the power law photon index and the high energy cut-off are directly linked to the temperature and optical depth of the Comptonising hot plasma. Knowing these two quantities can therefore provide vital information for understanding all the characteristics of the plasma near the central engine. The values of Γ and $E_c$ are in fact linked to the Comptonising hot plasma temperature kT$_e$ and optical depth $\tau$, according to the relation discussed by Petrucci et al. (2001):

$$\Gamma - 1 \simeq \left\{ \frac{9}{4} + \frac{m_e c^2}{kT_e \tau(1 + \tau/3)} \right\}^{1/2} - \frac{3}{2}$$

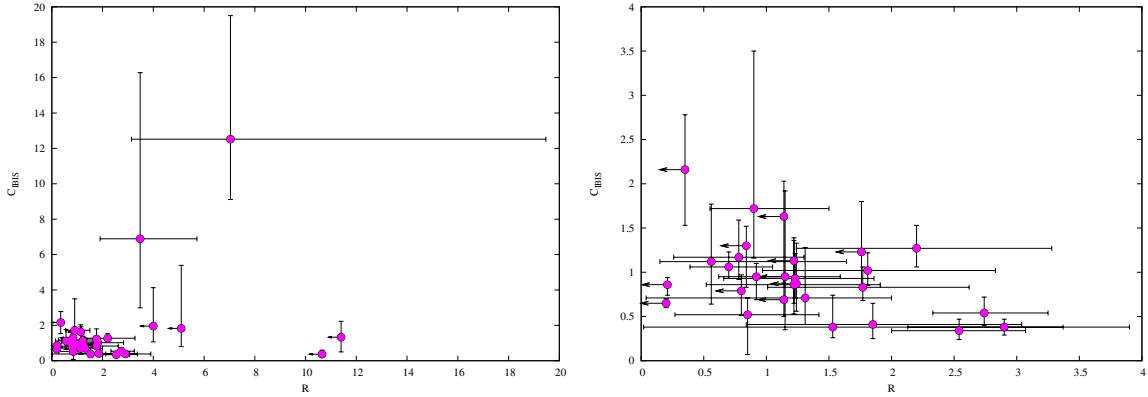

**Figure 13.** *Left Panel*: reflection fraction vs. cross-calibration constant between X-ray and IBIS data. *Right Panel*: a zoom of the plot for small values of *R*.

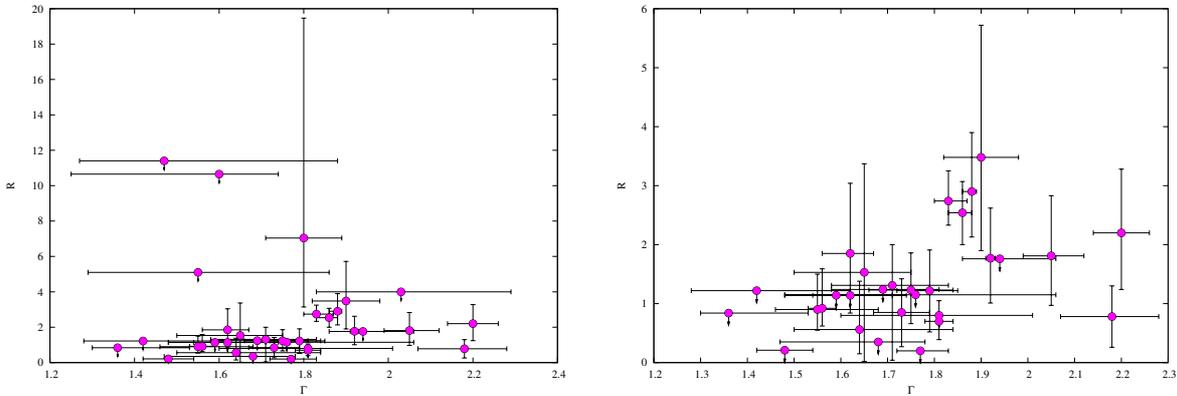

**Figure 14.** *Left Panel*: reflection fraction vs. photon index for the AGN in our sample. *Right Panel*: a zoom of the plot excluding sources with very high R values (see table 3).

The plasma temperature kT$_e$ is estimated as kT$_e$=E$_{cut}$/2 if $\tau \lesssim 1$ and kT$_e$=E$_c$/3 if $\tau \gg 1$. In our sample the mean value of E$_c$ is ~100 keV, with most objects in the sample falling in the range 50-150 keV; higher cut-off energies may be present but in a small number of AGN. The most likely range of E$_c$ estimated in the present sample, indicates a range of plasma temperatures from 20 to 80 keV (or 2-9×10$^8$ K). The equation (solved for both low and high values of $\tau$ and E$_c$ and assuming our average value of $\Gamma$=1.7) has therefore acceptable solutions for $\tau$ in the range 1 to 4. These results are in good agreement with what was previously found for a small sample of Seyfert 1 galaxies studied by Petrucci et al. (2001) and indicate that the plasma has a typical temperature of (50±30) keV and is not too thick ($\tau$<4). Available models should be able to explain and cover the observed range of values.

## 7 THE COMPLETE SAMPLE: COSMIC DIFFUSE X-RAY BACKGROUND

A fundamental input parameter for CXB synthesis models is the broad-band spectral shape of type 1 AGN. The main spectral component is, as we have found, a power law with an exponential roll over at high energies. The average slope of the power law component is generally taken to be $\Gamma$=1.9, mainly from observations performed at low energies or, over a broad band, by *BeppoSAX*; there seems to be no significant redshift dependence of $\Gamma$ up to high values of z. A harder average spectrum means a stronger contribution from unabsorbed AGN and hence a lower number of obscured sources required to match the CXB. Up to now the dispersion around this mean photon index value has generally been neglected ($\sigma$=0), but recently Gilli et al. (2007) have shown that the modeled CXB spectrum can be hardened near the 30 keV peak by as much as 20-30% by taking into account a dispersion in photon indices of 0.2 or more. This means that, besides the average $\Gamma$, also its dispersion becomes an important observational parameter to measure. Since observations above 10 keV have been less frequent up to now, due to the lack of instrument sensitivity, the cut-off energy is, together with the reflection bump, the parameter less well constrained. While it is well known that an exponential cut-off must be present around a few hundreds of keV, in order not to violate the present level of the CXB above 100 keV, values of 200 keV and up to 500 keV have been used in the past. This was mainly based on

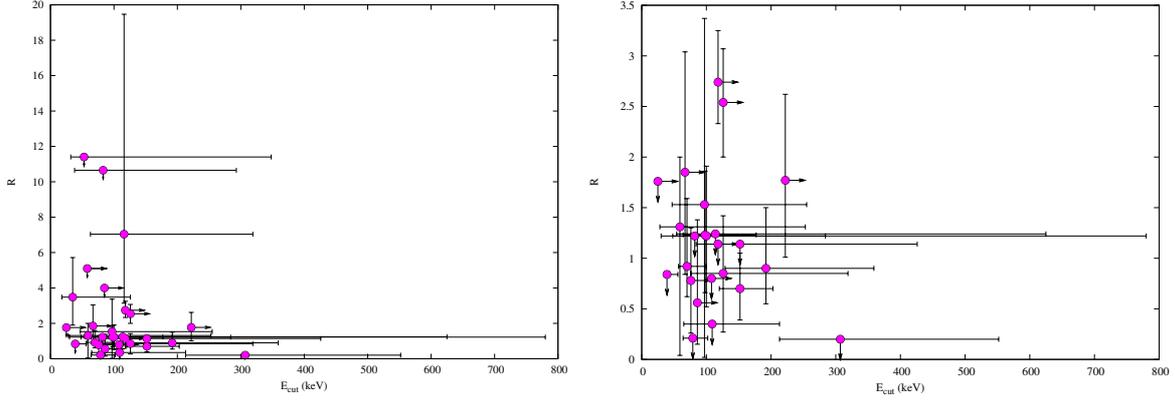

**Figure 15.** *Left Panel*: High energy cut-off vs. the reflection fraction for the complete sample of type 1 AGN. *Right Panel*: a zoom of the plot excluding sources with very high R values (see table 3).

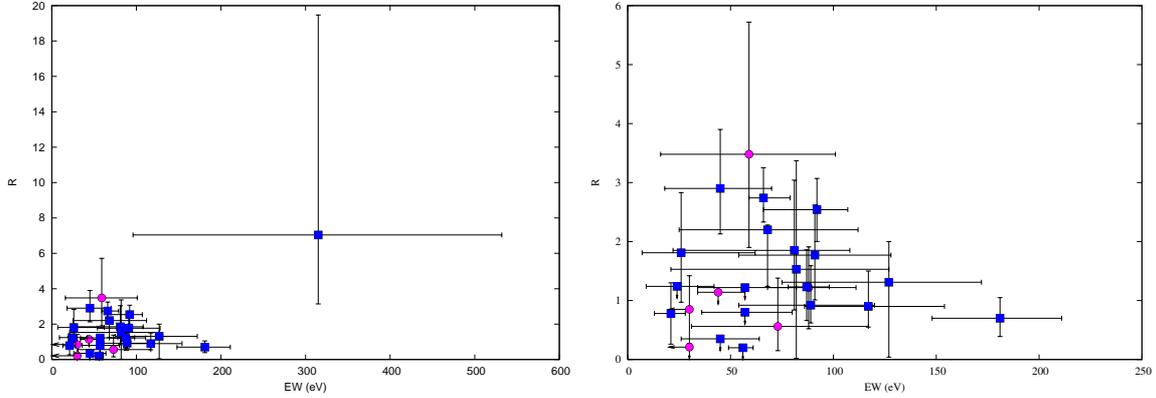

**Figure 17.** *Left Panel*: Reflection fraction vs. Equivalent Width for RL and RQ sources. *Right Panel*: a zoom of the plot in the region of low EW and R (NGC 6814 has been excluded). Magenta circles are radio loud sources, blue boxes are radio quiet objects.

*BeppoSAX* observations which indicate a relatively large spread in $E_c$ from about 50 keV up to about 300 keV or more (Perola et al. 2002; Dadina 2008). A better constraint on $E_c$ is therefore a very important result which modelers of CXB are waiting for.

The characteristic hard slope and the 40 keV break of the CXB indicate, besides the existence of heavily absorbed objects, the presence of at least some reflection in AGN spectral energy distributions (SEDs), and CXB synthesis models typically use a reflection fraction R ∼ 1 (Gilli et al. 2007; Gandhi et al. 2007). Clearly the higher the reflection, the less important becomes the contribution of Compton thick AGN, which are being found in small percentage in the local Universe both by *Swift* and *INTEGRAL* (Ajello et al. 2008; Bianchi et al. 2009).

With the above choices of parameters and carefully adjusting them by comparing the model predictions with the measured CXB spectrum, it is possible to reproduce a self-consistent picture, where the CXB is simply due to the integrated emission of absorbed and unabsorbed AGN; indeed the cumulative X-ray spectrum of local AGN selected at high energies by *Swift*/BAT and *INTEGRAL*/IBIS is able to replicate the shape of the CXB as demonstrated by Winter et al. (2009), Sazonov et al. (2008) and Triester et al. (2009). It is clear, however, that within the delicate interplay between parameters, a small variation of one of them can violate one or more of the observational constraints. A detailed modeling of the CXB using results reported here is beyond the scope of the present work, but clearly our data, which focused for the first time on a complete sample of type 1 AGN with detection at high energies, are able to provide key information on the AGN average continuum slope and dispersion and a better definition of the other two parameters used in CXB modeling (namely $R$ and $E_c$). We find that the primary continuum is a power law with slope ∼1.7 ($\sigma$=0.2) and a cut-off at around 110 keV ($\sigma$=60 keV) while the reflection fraction spans from 0.8 to 2.2. In the recent work of Gilli et al. (2007), a steeper power law ($\Gamma$=1.9), a higher cut-off energy (200 keV) and a reflection of 1 were assumed, while the spectral index dispersion was similar to the value found in the present work. A flatter photon index and a lower cut-off energy for the average spectrum of unabsorbed sources have clearly a strong impact on the synthesis model of the cosmic X-ray background, particularly with respect to the fraction of Compton thick sources required to match the 30 keV peak. Although our results are in contrast with previous studies, mainly from *BeppoSAX* data, we note that they agree with the recent estimate of the cumulative hard X-ray spectrum of local *INTEGRAL* AGN obtained by Sazonov et al. (2008): these authors find $\Gamma$=1.64±0.17 and $E_c$=124$^{+134}_{-38}$ keV, without taking into account the reflection component, which has the effect of steepening the spec-tral index to value closer to ours. Similarly, Winter et al. (2009), using a sample of *Swift/BAT* selected AGN, find that the average spectrum is very flat (∼1.4), again without taking into account reflection, and in this case also a high energy cut-off. Thus the observational evidence gathered so far by hard X-ray surveys points to a flatter average spectrum and possibly a lower cut-off as found in this work. It would be interesting to explore the consequences of fitting the CXB with the range of parameters found in the present work, particularly with reference to the fraction of Compton thick sources needed to provide a self-consistent fit; this clearly represents the next step of our work.

## 8 CONCLUSIONS

In this work we presented, for the first time, the broad-band spectral analysis of a complete sample of type 1 AGN, detected by *INTEGRAL* in the hard X-ray band (20-40 keV). This is a major step forward in the analysis of the spectral features of AGN, since previous studies mainly focused on small, non complete, samples of objects.

¿From a general point of view, the results presented in this work may be summarised as follows:

- Absorbing column densities are generally small or absent, except in those cases where complex absorption is required.
- The average photon index is 1.7 (dispersion of 0.2), flatter than the generally accepted canonical value of 1.9.
- The mean high energy cut-off is 110 keV with a small spread: most objects have $E_c$ in the range 50-150 keV; this average value is lower and the spread smaller than found in previous works based on broad-band spectral studies of type 1 galaxies.
- The average reflection fraction is close to 1.5 (dispersion of 0.7) only slightly higher than obtained in previous measurements and than assumed for synthesis models of the CXB.
- The iron lines detected in our sources tend to be narrow, with only 15% of the sample requiring a broad Fe line. The EW are generally below or around 100 eV, in general agreement with the measured reflection fractions.
- Several correlations in the parameter space have been investigated, but none is found, except for a weak correlation between the iron line equivalent width and the reflection fraction.
- The average cut-off energy and spectral index provide an estimate on the temperature and optical depth of the plasma where the hard X-ray emission originates. The temperature falls mostly in the 20 to 80 keV band (or 2-9$\times 10^8$ K) while the optical depth ranges from 1 to 4, implying that the Comptonising plasma is not not too thick.
- Finally, our result could provide a new self-consistent set of parameters for synthesis models of the Cosmic Diffuse X-ray Background.

## ACKNOWLEDGEMENTS

We acknowledge the University of Southampton financial support and financial contribution from contracts ASI-INAF I/016/07/0 and I/088/06/0 (M.M.) and the Italian Space Agency (ASI) financial and programmatic support via contract I/008/07/0. M.M. wishes to thank V.A. McBride (Southampton University) for reducing *Chandra* data for QSO B0241+62.

## REFERENCES

Ajello M., Rau A., Greiner J. et al. 2008, ApJ, 673, 96
Arnaud K. A. 1996, Astronomical Data Analysis Software and Systems V, eds Jacoby G. and Barnes J., p.17, ASP Conf Series vol. 101
Avni Y. & Bahcall J.N. 1980, ApJ, 235, 694
Barger A.J., Cowie L.L., Mushotzky R.F., Yang Y., Wang W.-H., Steffen A.T., Capak P. 2005, AJ, 129, 578
Baumgartner W., Tueller J., Mushotzky R. et al. 2008, ATel 1429
Beckmann V., Barthelmy S.D., Courvoisier T.J.-L., Gehrels N., Soldi S., Tueller J., Wendt, G. 2007, A&A, 475, 827
Bianchi S., Guainazzi M., Matt G. & Fonseca Bonilla N. 2007, A&A 467, L19
Bianchi S., Guainazzi M., Matt G., Fonseca Bonilla N. & Ponti G. 2009, A&A, 495, 421
Bird A. J., Malizia A., Bazzano A. et al. 2007, ApJS, 170, 175
Cappi M., Mihara T., Matsuoka M., Hayashida K., Weaver K.A., Otani C. 1996, ApJ, 458, 149
Comastri A., Setti G., Zamorani G. & Hasinger G. 1995, A&A, 296, 1
Comastri A., Gilli R. & Hasinger G. 2006 invited talk at the meeting "Gamma Wave 2005", Bonifacio, September 2005. To be published in "Experimental Astronomy"
Comastri A., Gilli R., Vignali C., Matt G., Fiore F., Iwasawa K. 2007, Progr. Theor. Phys. Suppl., 169, 274
Dadina M. 2007, A&A, 461, 1209
Dadina M. 2008, A&A, 485, 417
De Rosa A., Piro L., Fiore F., Grandi P., Maraschi L., Matt G., Nicastro F., Petrucci P.O. 2002, A&A, 387, 838
De Rosa A., Piro L., Perola G.C. et al. 2007, A&A, 463, 903
Elitzur M. & Shlosman I. 2006, ApJ, 648, 101
Fabian A.C., Vaughan S., Nandra K. et al. 2002, MNRAS, 335, 7
Fabian A.C., Miniutti G., Gallo L., Boller Th., Tanaka Y., Vaughan S., Ross R.R. 2004, MNRAS, 353, 1071
Fabian A.C., Miniutti G., Iwasawa K., & Ross R.R. 2005, MNRAS, 361, 795
Gandhi P., Fabian A.C., Suebsuwong T., Malzac J., Miniutti G., Wilman R.J. 2007, MNRAS, 382, 1005
Gilli R., Comastri A. & Hasinger G. 2007, A&A, 463, 79
Goldwurm A., David P., Foschini L. et al. 2003, A&A, 411, 223
Grandi P., Malaguti G. & Fiocchi M. 2006, ApJ, 642, 113
Gruber D.E., Matteson J.L., Peterson L. E. & Jung G. V. 1999, ApJ, 520, 124
Guainazzi M., Perola G.C., Matt G., Nicastro F., Bassani L., Fiore F., dal Fiume D., Piro L. 1999, A&A, 346, 407
Guainazzi M., Bianchi S. & Dovčiak M. 2006, AN, 327, 1032
Hickox R.C. & Markevitch M. 2006, ApJ, 645, 95
Hill J.E., Burrows D. N., Nousek J.A. et al. 2004, SPIE, 5165, 217
Landi R., De Rosa A., Dean A.J., Bassani L., Ubertini P., Bird A.J. 2007, MNRAS, 380, 926
Lebrun F., Leray J.P., Lavocat P. et al. 2003 A&A, 411, 141
Malizia A., Bassani L., Stephen J.B. et al. 2003, ApJ, 589, 17
Malizia A., Bassani L., Bird A.J et al. 2008, MNRAS, 389, 1360
Malzac J. 2001, MNRAS, 325, 1625
Malzac J., Beloborodov A. M. & Poutanen J. 2001, MNRAS, 326, 417
Malzac J. & Petrucci P.O. 2002, MNRAS, 336, 1209
Maraschi L. & Haardt F. 1997, ASPC, 121, 101
Markowitz A., Reeves J.N. & Braito V. 2006, ApJ, 646, 783
Masetti N., Landi R., Pretorius M.L. et al. 2007, A&A 470, 331
Matt G. 2001, AIPC, 599, 209

Matt G., Bianchi S., De Rosa A., Grandi P. & Perola G.C. 2006, A&A, 445, 451

Mattson B.J., Weaver K.A. and & Reynolds C.S. 2007 ApJ, 664, 101

Merloni A., Malzac J., Fabian A.C. & Ross R.R. 2006, MNRAS, 370, 1699

Miniutti, G. & Fabian A.C. 2004, MNRAS, 349, 1435

Miniutti G., Fabian A.C., Anabuki N. et al. 2007, PASJ, 59, 315

Molina M., Malizia A., Bassani L. et al. 2006, MNRAS, 371, 821

Molina M., Giroletti M., Malizia A. et al. 2007 MNRAS 382, 937

Molina M., Bassani L., Malizia A., Bird A.J., Dean A.J., Fiocchi M., Panessa F., de Rosa A. & Landi, R. 2008, MNRAS, 390, 1217

Mushtozky R. 1984, Adv. Space Res. Vol.3, N10-12, 157

Panessa F., Bassani L., De Rosa A. et al. 2008, A&A, 483, 151

Perola G.C., Matt G., Cappi M., Fiore F., Guainazzi M., Maraschi L., Petrucci P.O., Piro L. 2002, A&A, 389, 802

Petrucci P.O., Haardt F., Maraschi L. et al. 2001, ApJ, 556, 716

Piconcelli E., Jiménez-Bailón E., Guainazzi M., Schartel N., Rodríguez-Pascual P. M., Santos-Lleó M. 2005, A&A, 432, 15

Piro L. 1999, AN, 320, 236

Reeves, J. N. & Turner, M. J. L. 2000, MNRAS, 316, 234

Sambruna R.M., Eracleous M., Mushotzky R.F. 1999, ApJ, 526, 60

Sazonov S., Krivonos R., Revnivtsev M., Churazov E. & Sunyaev R. 2008, A&A, 482, 517

Schmidt M. 1968, ApJ, 151, 393

Spergel D.N., Verde L., Peiris H.V. et al. 2003, ApJS, 148, 175

Stephen J.B. et al. 2009, in preparation.

Strüder L., Briel U., Dennerl K. et al. 2001, A&A, 365L, 18

Triester E., Urry C.M., Virani S. 2009, ApJ, 696, 110

Tueller J. Mushotzky R.F., Barthelmy S. et al. 2008, ApJ, 681, 113

Turner M. J. L., Abbey A., Arnaud, M. et al. 2001, A&A, 365L, 27

Vaughan S. & Fabian A.C. 2004, MNRAS, 348, 1415

Winter L.M., Mushotzky R.F., Tueller J. & Markwardt C. 2008, ApJ, 674, 686

Winter L.M., Mushotzky R.F., Reynolds C.S. & Tueller J. 2009, ApJ, 690, 1322

Zdziarski A.A., Ghisellini G., George I.M., Fabian A. C. Svensson R., Done C. 1990, ApJ, 363, 1

Zdziarski A.A. Poutanen J., Mikolajewska J., Gierlinski M., Ebisawa K., Johnson W. N. 1998, MNRAS, 301, 435